\documentclass[twocolumn,showpacs,preprintnumbers,amsmath,amssymb]{revtex4}
\usepackage{tabularx,graphicx}

\usepackage{color}
\usepackage{hyperref}
\hypersetup{
    colorlinks=true,
    linkcolor=blue,
    filecolor=blue,      
    urlcolor=blue,
}

\begin{document}

\newcommand{\beq}{\begin{equation}}
\newcommand{\eeq}{\end{equation}}
\newcommand{\beqn}{\begin{eqnarray}}
\newcommand{\eeqn}{\end{eqnarray}}
\newcommand{\bmath}{\begin{subequations}}
\newcommand{\emath}{\end{subequations}}
\newcommand{\bra}[1]{\langle #1|}
\newcommand{\ket}[1]{|#1\rangle}

\title{Thermodynamic inconsistency of the conventional theory of superconductivity}
\author{J. E. Hirsch }
\address{Department of Physics, University of California, San Diego,
La Jolla, CA 92093-0319}

\begin{abstract} 
A type I superconductor expels a magnetic field from its interior  to a surface layer of thickness $\lambda_L$, the London penetration depth.
$\lambda_L$ is a function of temperature, becoming smaller as the temperature decreases. Here we analyze the process of cooling (or heating)
a type I superconductor in a magnetic field, with the system remaining always  in the superconducting state. The conventional theory  predicts that Joule heat is generated in this process, the amount of which depends on the rate at which the temperature changes. 
Assuming the final state of the superconductor is independent of history, as the conventional theory assumes, we show that this process violates the first and second laws of thermodynamics. We conclude that the conventional theory of superconductivity is internally inconsistent. Instead, we suggest that the
alternative theory of hole superconductivity may be able to resolve this problem.
 \end{abstract}
\maketitle

           \begin{figure} [t]
 \resizebox{7.5cm}{!}{\includegraphics[width=6cm]{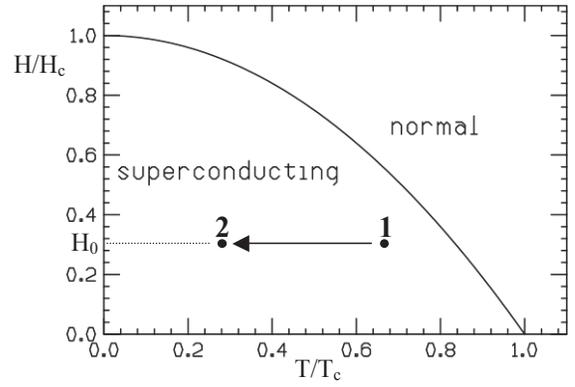}} 
 \caption { Critical magnetic field versus temperature for a type I superconductor. We will consider the process where a system evolves from point 1 to point 2 along
 the direction of the arrow.  }
 \label{figure1}
 \end{figure} 
 
           \begin{figure} 
 \resizebox{8.4cm}{!}{\includegraphics[width=6cm]{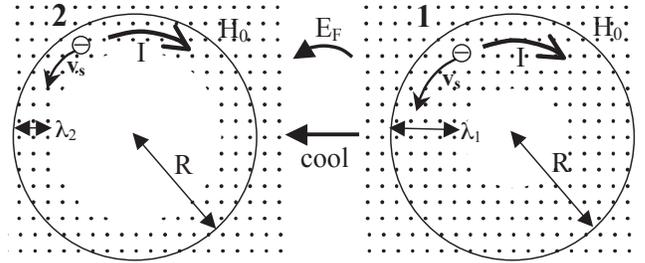}} 
 \caption { Cylindrical superconductor seen from the top. The right (left) panel indicates the system in the state 1 (2) of Fig. 1. 
 The dots indicate magnetic field $H_0$ coming out of the paper. The same current $I$ flows
 in both states.  The Faraday electric field $E_F$ generated during the process points counterclockwise.}
 \label{figure1}
 \end{figure} 

\section{Introduction}

After the discovery of the Meissner effect, it was concluded that the superconducting state of a simply connected body in the presence of a magnetic field lower
than the critical field is a thermodynamic state of matter and not 
a metastable  non-equilibrium state that depends on history, as was previously believed \cite{gortercasimir}. Conventional superconductors are believed to be
described by London theory and BCS theory, which we will call the conventional theory \cite{tinkham}. In this paper   we show that the conventional theory cannot
describe certain processes in type I superconductors without violating well established laws of physics. 

Figure 1 shows the phase diagram of a type I superconductor in a magnetic field $H$. We consider the process where a cylindrical superconductor 
is cooled from state 1 to state 2 shown in Fig. 1, in the presence of an applied field $H_0$. Figure 2 shows the superconductor as seen from the top, with the dots
indicating magnetic field pointing out of the paper. We will discuss the details in the following. We will conclude that  the conventional theory is
  inconsistent with the laws of thermodynamics.

\section{basic equations in simplest form}

We consider a long cylinder of radius $R$ and height $h$, with $R<<h$, in an applied uniform magnetic field $H_0$ parallel to its axis. 
Assume the cylinder is in the superconducting state, hence the magnetic field in the deep interior is zero. An azimuthal current circulates near its surface.
From Ampere's law we have 
\beq
\oint \vec{B}\cdot \vec{d\ell}=\frac{4\pi}{c}\int \vec{j}_s\cdot \vec{dS}=\frac{4\pi}{c}I
\eeq
where $\vec{j}_s$ is the current density and $I$ is the total current.
Taking one side of the contour in the deep interior and the other one outside the cylinder we obtain
\beq
I=\frac{c}{4\pi}hH_0
\eeq
Assuming the current circulates within a layer of thickness $\lambda_L$  of the surface, with current density $j_s$, the  total current and current density 
are related by 
\beq
I=j_s\lambda_L h
\eeq
hence the current density is 
\beq
j_s=\frac{c}{4\pi \lambda_L}H_0 .
\eeq
The London penetration depth varies with temperature, becoming smaller as the temperature decreases. 
From Eq. (4) we see that the supercurrent density increases as the temperature is lowered, but since it flows in an 
increasingly thinner layer the total current Eq. (2) is independent of temperature. 

The current density is related to the superfluid density $n_s$ and the superfluid velocity $\vec{v}_s$ by
\beq
\vec{j}_s=n_s e \vec{v}_s
\eeq
with $e$ the electron charge, and within London-BCS theory the superfluid velocity is given by \cite{tinkham}
\beq
\vec{v}_s=-\frac{e}{m_ec}\vec{A}
\eeq
where $\vec{A}$ is the magnetic vector potential. For simplicity we ignore any difference between bare mass $m_e$ and 
effective mass \cite{mstar}, this will not affect our results. From Eqs. (5) and (6)
\beq
\vec{j}_s=-\frac{n_s e^2}{m_e c}\vec{A}
\eeq
so that 
\beq
\vec{\nabla}\times\vec{j}_s=-\frac{n_s e^2}{m_e c}\vec{B}
\eeq
with $\vec{B}=\vec{\nabla}\times \vec{A}$.  From  Ampere's law in differential form 
\beq
\vec{\nabla}\times\vec{B}=\frac{4\pi}{c}\vec{j}_s
\eeq
taking the curl on both sides and using Eq. (8)
\beq
\nabla^2\vec{B}=\frac{4\pi n_s e^2}{m_e c^2}\vec{B}=\frac{1}{\lambda_L^2}\vec{B} .
\eeq
so that the magnetic field and the supercurrent decay to zero exponentially in the interior over distance  $\lambda_L$,
given in Eq. (10) as a function of superfluid density $n_s$. From Eq. (10) 
\beq
n_s \lambda_L^2=\frac{m_e c^2}{4\pi e^2}  .
\eeq
As the temperature is lowered, $n_s$ increases and $\lambda_L$ decreases keeping the product in Eq. (11) constant.  
BCS theory provides equations that give the temperature dependence of $\lambda_L$. The BCS prediction is very close to
the behavior predicted by the two-fluid model \cite{gortercasimir}
 \beq
 \frac{1}{\lambda_L(T)^2}= \frac{1}{\lambda_L(0)^2}(1-(\frac{T}{T_c})^4)
 \eeq
 except at very low temperatures where BCS predicts exponential rather than power-law temperature dependence \cite{tinkham}. 
 However we will not be interested in that regime, and will use Eq. (12) in this paper for simplicity.

From Eq. (6),
using Stokes' theorem and the fact that the supercurrent flows in the surface layer of thickness $\lambda_L$ 
we obtain for the magnitude of the superfluid velocity
\beq
v_s=-\frac{e\lambda_L}{m_e c} B
\eeq
so the superfluid velocity decreases as the temperature is lowered and $\lambda_L$ decreases. The magnitude of the current density in terms of the magnetic field is given by
\beq
 j_s=\frac{n_s e^2}{m_e c}\lambda_L  B=\frac{c}{4\pi \lambda_L} B .
\eeq

We will also be interested in the mechanical angular momentum carried by the electrons in the supercurrent. The mechanical momentum density of electrons at position $\vec{r}$ is \cite{rostoker}
\beq
\vec{\mathcal{P}}(\vec{r})=\frac{m_e}{e}\vec{j}_s(\vec{r})
\eeq
where $\vec{j}_s(\vec{r})$ is the current density at position $\vec{r}$. Assuming the supercurrent circulates on the surface layer of thickness $\lambda_L$
  and using Eqs. (5), (11) and (13)    the total mechanical angular momentum of the electrons in the supercurrent is, assuming $\lambda_L<<R$ 
\beq
\vec{L}_e=-\frac{m_e c}{2e}hR^2 H \hat{z}  
\eeq
so that the total electronic angular momentum does not change with temperature. Consequently we can conclude by momentum conservation that
the body as a whole does not change its angular momentum when the system is cooled.

\section{the process and the questions}
We consider the process shown in Fig. 1, where a superconducting cylinder in a magnetic field is cooled from initial temperature $T_1$ to
finite temperature $T_2$. The London penetration depth changes from $\lambda_L(T_1)\equiv \lambda_1$ to $\lambda_L(T_2)\equiv \lambda_2$,
with $\lambda_2<\lambda_1$.

During this process, magnetic field is expelled, since  initially it penetrates up to radius $r_1\sim R-\lambda_1$ and at the end only to radius
$r_2\sim R-\lambda_2$, as shown schematically  in Fig. 2. In other words, the field was expelled from the region $R-\lambda_1<r<R-\lambda_2$, a ring of thickness
$\lambda_1-\lambda_2$. This gives rise to a Faraday electric field $E_F$ pointing in counterclockwise direction that is determined by Faraday's law,
\beq
\oint \vec{E}_F\cdot \vec{d\ell}=-\frac{1}{c}\frac{\partial}{\partial t}\int \vec{B}\cdot\vec{dS}
\eeq
If the process occurs over a time $t_0$ Eq. (17) yields  for the time integral of the Faraday field
\beq
\int_0^{t_0}E_F dt=\frac{H_0}{c}(\lambda_1-\lambda_2)
\eeq
The Faraday field acts on the superfluid electrons as
\beq
\frac{d}{dt}v_s=\frac{e}{m_e}E_F
\eeq
and integrating both sides of Eq. (19) over time and using Eq. (18), the change in superfluid velocity is
\beq
\Delta v_s=\frac{e}{m_e}\int_0^{t_0}E_F dt=\frac{eH_0}{m_ec}(\lambda_1-\lambda_2)=\frac{eH_0}{m_ec}\Delta \lambda_L
\eeq
which is precisely what Eq. (13) predicts. So as the London penetration depth decreases from $\lambda_1$ to $\lambda_2$ the superfluid velocity decreases 
as given by Eq. (13), and the time variation of the superfluid velocity is completely accounted for by the action of the Faraday electric field
on the superfluid carriers.

According to the above, the Faraday field, pointing in counterclockwise direction,  will change the angular momentum of the superfluid electrons, increasing their momentum in clockwise direction. At the same time, 
it will impart counterclockwise angular momentum to the positive ions, i.e. to the body as a whole. However we have seen that the total electronic angular momentum 
$L_e$ (Eq. (16)) doesn't change in this process, hence neither does the total ionic angular momentum. How is this possible?

The reason is, as the temperature is lowered and $\lambda_L$ decreases  the number of superfluid electrons $n_s(T)$ will increase, according to Eq. (11).
In the process of normal electrons condensing into the superconducting state they have to  acquire counterclockwise momentum, of exactly the right magnitude to cancel the
clockwise momentum imparted on the superfluid electrons by the Faraday field. And the same process  of condensation
has to impart  clockwise momentum  to the body as a whole
to counteract the counterclockwise momentum imparted to the body by the Faraday field.

The theory of superconductivity has to explain the physical mechanisms by which these changes in momenta happen in the process of normal electrons condensing into the
superconducting state. To the best of our knowledge this has never been discussed in the superconductivity literature. 
In refs \cite{momentum,entropy} we argued that BCS-London theory does not have the physical elements necessary to explain these processes.

Furthermore, we have to remember that at any given temperature there are both superfluid and normal electrons, of density $n_s$ and $n_n$, with $n_s+n_n=n$ constant in time, in a two-fluid description. Similarly within BCS theory there is the superfluid and Bogoliubov quasiparticles at finite
temperature, we will call the latter `normal electrons'.
The Faraday electric field will impart clockwise momentum to these normal electrons, and this momentum will decay to zero through scattering with impurities or phonons.
These are irreversible processes, that generate Joule heat and entropy. However, it should be possible to cool a thermodynamic system in a reversible way, without
entropy generation.

In the following sections we analyze the dynamics and the thermodynamics of these processes   in  detail and conclude that 
they cannot be understood within the 
  conventional theory of superconductivity, and for that reason the conventional  theory is incompatible with the laws of physics.

\section{exact equilibrium equations for the cylinder}
We recall briefly the exact solution of the equilibrium electrodynamic equations for an infinite cylinder \cite{laue}. Equation (10)  for the magnetic field $\vec{B}$ is, in cylindrical coordinates
\beq
\frac{1}{r}\frac{\partial }{\partial r} (r\frac{\partial B(r)}{\partial r})-\frac{1}{\lambda_L^2}B(r)=0
\eeq
assuming cylindrical symmetry. From Eq. (9), the current is given by
\beq
\vec{j}_s(r)=-\frac{c}{4\pi} \frac{\partial B}{\partial r} \hat{\theta}
\eeq
With boundary condition $B(R)=H_0$ the solution is
\bmath
\beq
B(r)=H_0\frac{J_0(ir/\lambda_L)}{J_0(iR/\lambda_L)}
\eeq
\beq
j_s(r)=\frac{c}{4\pi\lambda_L}H_0i\frac{J_1(ir/\lambda_L)}{J_0(iR/\lambda_L)}
\eeq
\emath
where $J_0$ and $J_1$ are Bessel functions, related by $J_1(x)=-(\partial/\partial x) J_0(x)$. They are given by the series expansions
\bmath
\beq
J_0(x)=\sum_{m=0}^\infty (-1)^m\frac{(\frac{1}{2}x)^{2m}}{(m!)^2}
\eeq
\beq
J_1(x)=\sum_{m=0}^\infty (-1)^m\frac{(\frac{1}{2}x)^{2m+1}}{(m!)(m+1)!}
\eeq
\emath
and for large imaginary argument by
\bmath
\beq
J_0(ix)= \frac{e^x}{\sqrt{x}}
\eeq
\beq
J_1(ix)=i \frac{e^x}{\sqrt{x}} .
\eeq
\emath
In the following we assume $R>>\lambda_L$ and consider all quantities derived from these equations to lowest order in $(\lambda_L/R)$ only, for simplicity.
We don't think this affects our reasoning and conclusions. In any event, it is straightforward to extend the treatment to avoid this approximation.

\section{ equations for large $R/\lambda_L$}
We consider a process where the temperature changes gradually so the system is always in equilibrium, between an initial time $t=0$ and a final time $t=t_0$.
The London penetration depth is given by $\lambda_L(t)$, with $\lambda_L(0)=\lambda_1$, $\lambda_L(t_0)=\lambda_2$. The magnetic field is given by
\beq
\vec{B}(r,t)=H_0 e^{(r-R)/\lambda_L(t)} \hat{z}
\eeq
the magnetic vector potential by
\beq
\vec{A}(r,t)= H_0 \lambda_L(t)e^{(r-R)/\lambda_L(t)} \hat{\theta}
\eeq
the Faraday electric field by
\beqn
\vec{E}(r,t)&=& -\frac{1}{c}\frac{\partial A(r,t)}{\partial t} \\  \nonumber 
&=&
-\frac{H_0}{c}(1+\frac{R-r}{\lambda_L})e^{(r-R)/\lambda_L(t)}\frac{\partial \lambda_L}{\partial t}\hat{\theta}
\eeqn
the current density by
\beq
\vec{j}_s(r,t)=-\frac{c}{4\pi \lambda_L} H_0 e^{(r-R)/\lambda_L(t)} \hat{\theta}
\eeq
 the superfluid velocity by
\beq
\vec{v}_s(r,t)=-\frac{eH_0}{m_e c}\lambda_L(t)e^{(r-R)/\lambda_L(t)} \hat{\theta}
\eeq
and the London penetration depth and superfluid density satisfy
\beq
n_s(t) \lambda_L(t)^2=\frac{m_e c^2}{4\pi e^2}  .
\eeq

Note that, from Eqs. (28) and (30)
\beq
\frac{d\vec{v}_s(r,t)}{dt}=\frac{e}{m_e}\vec{E}(r,t) ,
\eeq
in other words, the Faraday electric field causes the superfluid velocity to change (slow down) according to
Newton's law, as one would expect.

The total electronic angular momentum at time $t$ due to superfluid electrons only  is given by (we omit time dependence for brevity)
\beq
\vec{L}_e^s(t)=hn_s \int d^2r m_e v_s(r) r =2\pi hn_s m_e \int_0^R dr r^2 v_s(r)
\eeq
and we know from Eq. (16)  that $dL_e/dt=0$, i.e. it is a constant of motion. We can write
\beq
\frac{dL_e}{dt}=\frac{dL_e^{(1)}}{dt}+\frac{dL_e^{(2)}}{dt}
\eeq
with
\beqn
\frac{dL_e^{(1)}}{dt}&=&2\pi hn_s  \int_0^R dr r^2m_e \frac{d v_s(r)}{dt} \\ \nonumber
&=&2\pi hn_s  \int_0^R dr r^2 eE(r,t)   
\eeqn
where we used Eq. (32). 
This term has negative sign and expresses the fact that the Faraday electric field decreases the angular momentum of electrons in the supercurrent
in the process where the temperature is lowered and the London penetration depth decreases. 
This is because the Faraday field wants to restore the magnetic field in the interior that is being pushed further out in this process, by reducing
the supercurrent. But we know from Eq. (2) that the total supercurrent $I$ doesn't change.

The second term in Eq. (34) then has positive sign, it increases the electronic angular momentum. It is given by (to lowest order in $\lambda_L/R$)
\beq
\frac{dL_e^{(2)}}{dt}=\frac{\partial n_s}{\partial t} [2\pi R \lambda_L h] m_e v_s(R)R  .
\eeq
The term in square brackets is the volume of the surface layer of thickness $\lambda_L$ . What Eq. (36) says is that 
as electrons that are near the surface condense from the normal into the superconducting state, they `spontaneously' acquire the speed $v_s(r)$ and the electronic angular momentum changes accordingly.
Of course electrons in the interior also condense, but because they don't carry current they don't contribute to the change in angular momentum.

Associated with these electronic angular momentum changes there are also corresponding changes in the angular momentum of the ions, i.e. the body as a whole,
\beq
\frac{dL_i}{dt}=\frac{dL_i^{(1)}}{dt}+\frac{dL_i^{(2)}}{dt}=0
\eeq
with 
\bmath
\beq
\frac{dL_i^{(1)}}{dt}=-\frac{dL_e^{(1)}}{dt}
\eeq
\beq
\frac{dL_i^{(2)}}{dt}=-\frac{dL_e^{(2)}}{dt} .
\eeq
\emath
The first one, Eq. (38a),  is easily understood. Just like the Faraday electric field transfers clockwise momentum to the electrons, it transfers equal in magnitude and opposite in direction (i.e. counterclockwise)
momentum to the positive ions.
The second one, Eq.  (38b),  is clockwise momentum transferred to the body when normal electrons near the surface condense into the superconducting state.

The theoretical explanation of these processes has to explain the physical mechanism(s) that cause(s) normal electrons to change their angular momentum when they condense into the superconducting state, Eq. (36), 
and at the same time cause(s) the body to acquire the same angular momentum in opposite direction. These questions have not yet been discussed in the literature on  conventional  
superconductivity. We hope this will be done soon. We don't believe it is possible to understand these processes within the conventional
theory \cite{momentum,entropy}.

\section{normal current and joule heat}
Within London theory we can think of the superconductor as a two-fluid model, a mixture of normal fluid and superfluid, with proportions that vary with temperature.
The same is true within BCS theory, where the normal fluid is composed of Bogoliubov quasiparticles. 
The normal fluid will be subject to normal scattering processes. In the presence of an  electric field, a normal current will be generated, and dissipation
will occur. This is clearly illustrated by the behavior of superconductors with ac currents, as discussed e.g. in
ref. \cite{tinkhamac}.

The normal current flowing in counterclockwise direction induced by the electric field  is given by
 \beq
 \vec{j}_n(r,t)=\sigma_n(t) \vec{E}(r,t)
 \eeq
 with
 $\sigma_n$ the normal conductivity, which we can write as
 \beq
 \sigma_n(t)=\frac{n_n(t)e^2}{m_e}\tau
 \eeq
 where $\tau$ is the Drude scattering time and where  the normal electron density is given by
\beq
n_n(t)=n-n_s(t) 
\eeq
with  $n_s(t)$  given by Eq. (31), and $n$ the superfluid density at zero temperature, given by
\beq
n \lambda_L^2(T=0)=\frac{m_e c^2}{4\pi e^2}  .
\eeq
The power dissipated per unit volume is
\beq
\frac{\partial w}{\partial t}=\sigma_n(t) E(r,t)^2
\eeq
with the electric field given by Eq. (28). Integrating Eq. (43) over the   volume of the cylinder  we find for the Joule heat dissipated per unit time
\beq
 \frac{\partial W}{\partial t}\equiv \int d^3r \frac{\partial w}{\partial t}
 =\sigma_n\frac{H_0^2}{c^2}(\frac{\partial \lambda_L}{\partial t})^2 \frac{\pi h R \lambda_L(t)}{2}
 \eeq
 We can write the normal state conductivity in the form
\beq
\sigma_n(t)=\frac{n_n(t)}{n_s(t)}\frac{1}{\lambda_L(t)^2}\frac{c^2}{4\pi} \tau =(1-\frac{\lambda_L(0)^2}{\lambda_L(t)^2})\frac{1}{\lambda_L(0)^2}\frac{c^2}{4\pi} \tau 
\eeq
so Eq. (44) is
\beq
 \frac{\partial W}{\partial t} 
 = \frac{H_0^2}{8\pi } (\frac{1}{\lambda_L(0)^2}  -\frac{1}{\lambda_L(t)^2})     ( \frac{\partial \lambda_L}{\partial t})^2  (\pi h R \lambda_L(t) )\tau .
 \eeq
 Note that the total Joule heat dissipated, i.e. the time integral of Eq. (46), will depend on the speed of the process. For example, if we assume  that $\partial \lambda_L/\partial t$ is time-independent, we have
 \beq
\int _0^\infty  \frac{\partial W}{\partial t} dt =
  (\frac{\partial \lambda_L}{\partial t})
 \int _{\lambda_1}^{\lambda_2} d\lambda \frac{H_0^2}{8\pi } (\frac{1}{\lambda_L(0)^2} -\frac{1}{\lambda ^2})       (\pi h R \lambda )\tau 
 \eeq
 so the total Joule heat generated is proportional to the rate of change of the London penetration depth with time.
 
 Near the critical temperature, the London penetration depth varies extremely rapidly with temperature and it is clear that Eq. (46) can become very large.
 Since in the laboratory it is simpler to control the change in temperature with time rather than  the London penetration depth, let us rewrite Eq. (46) in terms
 of the former assuming for simplicity the  relation derived from the two-fluid model
 \beq
 \frac{1}{\lambda_L(t)^2}= \frac{1}{\lambda_L(0)^2}(1-(\frac{T(t)}{T_c})^4)
 \eeq
so we can write, in terms of $T=T(t)$, 
 \beq
 (\frac{\partial \lambda_L}{\partial t})^2=4(\frac{T}{T_c})^6\frac{\lambda_L(0)^2}{[1-(\frac{T}{T_c})^4]^3}\frac{1}{T_c^2}(\frac{\partial T}{\partial t})^2
 \eeq
 and Eq. (46) is
 \beqn
 \frac{\partial W}{\partial t} 
 &=& \frac{H_0^2}{8\pi }   (\frac{T}{T_c})^{10} \\ \nonumber
 &\times & \frac{1}{[1-(\frac{T}{T_c})^{4}]^{7/2}}\frac{1}{T_c^2}(\frac{\partial T}{\partial t})^2 (4\pi h R \lambda_L(0) )\tau .
 \eeqn
    Equation (50) says that if we cool a superconductor below $T_c$ at any reasonable cooling rate,
 the amount of Joule heat  dissipated will be arbitrarily large provided we start the cooling process sufficiently close to $T_c$.
 The same will be true if we heat a superconductor below $T_c$ to a temperature very close to $T_c$.
 
 However we also need to consider that  in a finite magnetic field $H_0$, the divergence in the London penetration depth as $T$ approaches $T_c$ will be cut off because the system will
 undergo a first order transition to the normal state. Nevertheless, the problem persists. Let us assume for simplicity   the two-fluid model expression for the
 critical magnetic field as function of temperature
 \beq
 H_0=H_c(1-(\frac{T_1}{T_c})^2)
 \eeq
 where $H_c$ is the zero temperature critical field, and $T_1$ is a temperature close to $T_c$:
 \beq
\frac{T_1}{T_c}=1-\delta_1
\eeq
with $\delta_1$ small. For the system at temperature $T$ below $T_1$, given by
\beq
\frac{T}{T_c}=1-\delta_1-\delta_2
\eeq
 we have for Eq. (50)
  \beqn
 \frac{\partial W}{\partial t} 
 &=& \frac{H_c^2}{8\pi }  (\frac{T}{T_c})^{10} \\ \nonumber
 &\times & 
 \frac{1}{8\delta_1^{3/2}(1+\delta_2/\delta_1)^{7/2}}
  \frac{1}{T_c^2}(\frac{\partial T}{\partial t})^2 (4\pi h R \lambda_L(0) )\tau .
 \eeqn
 which becomes arbitrarily large for sufficiently small values of $\delta_1, \delta_2$.

 The reader may argue that the divergence that we are pointing out is not really a problem. 
 First, because of the smallness of $\tau$, for any realistic values of $\partial T/\partial t$ Eq. (54) will be extremely small unless we are extremely close
 to $T_c$. The reader may argue that in that unrealistic regime some other physics may come in, involving fluctuations, that invalidates our simple treatment.
 
 However, we argue that the Joule-heat dissipation pointed out here brings other fundamental problems beyond the quantitative divergence pointed out
 above. Superconductors are supposed to be governed by conventional thermodynamics, and it should be possible to describe $reversible$ cooling and
 heating in the superconducting state. The generation of normal current due to the Faraday field and associated dissipation makes this impossible.
 In the next sections we show explicitly that thermodynamics is violated.

 \section{more normal current}
 
 There is in fact another contribution to the normal current besides the one given by Eq. (39) according to the conventional theory.
 
As the temperature decreases   the superfluid density increases, by additional normal
electrons joining the superfluid.  Condensing  normal electrons    `spontaneously' acquire the   velocity of electrons in the supercurrent, thus
increasing the electronic  momentum in counterclockwise direction according to Eq. (36). Conservation of momentum requires that the ions acquire equal momentum in opposite direction, i.e. clockwise, Eq. (38b). How does that happen? 
  
According to the conventional theory, as normal electrons condense 
a clockwise momentum imbalance is created in the normal region  \cite{halperin, entropy}, which is transferred to the body by normal collisions.
This momentum imbalance  corresponds to a $counterclockwise$ normal current that adds to the one induced by the Faraday field considered
in  Sect. VI. So the Joule heat generated will be  even larger than computed in Sect. VI.

Quantitatively, the change in superfluid density for a given change in penetration depth is, from Eq. (11)
 \beq
 \Delta n_s=-2n_s \lambda_L \Delta \lambda_L
 \eeq
 and the change in normal electron density is $\Delta n_n = -\Delta n_s$.   In a time interval $\tau$, the normal collision time, the normal electron density then changes by
  \beq
 \Delta n_n= 2n_s \lambda_L \frac{\partial \lambda_L}{\partial t} \tau  
 \eeq
The change in normal electron momentum density 
 in that  time interval  is
 \beq
\Delta \mathcal{P}=\Delta n_n m_e v_s
\eeq
since each condensing normal electron aquired velocity $v_s$, and the resulting normal current density is
\beq
j'_n(r,t)=\frac{e}{m_e} \Delta \mathcal{P}=2 \frac{n_s e^2}{m_e}   \tau  \frac{H_0}{c} e^{(r-R)/\lambda_L}  \frac{\partial \lambda_L}{\partial t} 
\eeq
where we used Eq. (30) for $v_s$. The Joule heat dissipated per unit time due to this current in the presence of the electric field
$E(r,t)$, Eq. (28),  is
\beq
\frac{\partial W'} {\partial t}=\int d^3r j'_n(r,t) E(r,t)
\eeq
and yields
\beq
\frac{\partial W'} {\partial t}=2\frac{n_s}{n_n}\sigma_n \frac{H_0^2}{c^2}(\frac{\partial \lambda_L}{\partial t})^2 \frac{\pi h R \lambda_L(t)}{2}
\eeq
and added to Eq. (44) yields for the total Joule heat per unit time
\beq
 \frac{\partial W_{total}} {\partial t}=[1+2\frac{n_s(t)}{n_n(t)}]\frac{\partial W} {\partial t}
 \eeq
 where $\partial W/\partial t$ was calculated in the previous section. So this new term is larger than the one calculated in the previous section at low temperatures, and smaller at high temperatures.
 In the two-fluid model, the crossover point is at  $T=0.904T_c$. So we conclude that over most of the temperature range below $T_c$, 
 the Joule heat due to the normal current generated in the condensation process dominates over the one due to the normal current induced by the  Faraday field.
 
               \begin{figure} [b]
 \resizebox{7.5cm}{!}{\includegraphics[width=6cm]{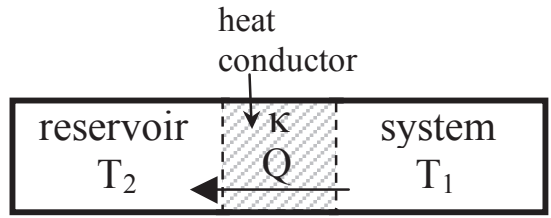}} 
 \caption { The system (superconductor in a magnetic field) at initial  temperature $T_1$ is connected to a heat reservoir at
 temperature $T_2<T_1$ through a heat conductor of thermal conductivity $\kappa$. The entire assembly is thermally and 
 mechanically insulated from its environment.    }
 \label{figure1}
 \end{figure}

 \section{thermodynamic inconsistency}
 Irrespective of how the normal current originated, we argue that its presence and the resulting Joule heat is in contradiction with
 thermodynamics. 
 
 Consider the situation shown in Fig. 3. The system is our superconductor with phase diagram given in figure 1, with applied magnetic field
 $H_0$. 
 The system is initially in thermal equilibrium at temperature $T_1$, with London penetration depth $\lambda_L(T_1)$.
 
 We put it in thermal contact with a heat reservoir at temperature $T_2<T_1$ through a thermal conductor with thermal conductivity
 $\kappa$. Heat will flow and eventually the system will reach temperature $T_2$ and be in thermal equilibrium with the heat
 reservoir. We assume the entire assembly  is thermally and mechanically insulated from its environment.
 The magnetic field originates in external permanent magnets, no work is performed on those magnets during the process. 
 We also assume the process is sufficiently slow that no electromagnetic radiation is generated.

 Given the initial and final states, we can compute various thermodynamic quantities. The total heat $Q$ transferred from the system to
 the reservoir during the process is
 \beq
 Q=\int_{T_2}^{T_1} dT C(T)
 \eeq
 where $C(T)$ is the heat capacity of our system. The change in entropy of the system in this process is
 \beq
 \Delta S=S(T_2)-S(T_1)=\ \int_{T_1}^{T_2} dT  \frac{C(T)}{T}
 \eeq 
 and is of course negative since $T_2<T_1$. The   change in entropy of the universe in this process is
 \beq
 \Delta S_{univ}=\frac{Q}{T_2}+\Delta S
 \eeq
and  is of course positive since we are dealing with an irreversible process, heat conduction between systems at different temperatures.
 
 The Joule heat generated in this process can be calculated from time integration of Eq. (61)
 \beqn
Q_J&=&\int_0^\infty  dt \frac{\partial W_{total}}{\partial t}=   \frac{H_0^2}{8\pi }  \int_0^\infty  dt 
 (1+2\frac{n_s(t)}{n_n(t)})    \\ \nonumber &\times&  (1-\frac{\lambda_L(0)^2}{\lambda_L(t)^2})
\frac{1}{\lambda_L(0)^2}  
   ( \frac{\partial \lambda_L}{\partial t})^2  (\pi h R \lambda_L(t) )\tau .
 \eeqn
 In addition, the Joule heat will generate entropy, given by
 \beq
 S_J=\int_0^\infty  dt  \frac{\partial W_{total}}{\partial t}\frac{1}{T(t)} .
 \eeq
 It is clear that the magnitudes of $Q_J$ and $S_J$ will depend on how fast the system is evolving from the initial to the final state,
 being larger for larger $\partial \lambda_L/\partial t$,
 which in turn will depend on the thermal conductivity of the heat  conductor, $\kappa$,  that connects the system with the heat reservoir.
 If  $\kappa$ is extremely low, essentially no Joule heat will be  generated nor Joule entropy.
 If $\kappa$  is not extremely low, these quantities will not be negligible.
 
 However, this does not make sense. Our system and the heat reservoir constitute our universe, their energy and entropy  are functions of state, and the initial
 and final states   in our process for both the system and the reservoir are uniquely defined. Therefore  the heat transferred  $Q$ and the change in entropy of the universe $\Delta S_{univ}$ are uniquely
 defined by Eqs. (62)-(64). There is no room for either $Q_J$ nor $S_J$.
 But  within the conventional theory of superconductivity, a normal current is necessarily generated when the 
 temperature changes below $T_c$, and nonzero Joule heat and Joule entropy are necessarily generated, unless the
 process happens infinitely slowly.
 
 It is important to emphasize that this argument   does not depend on the reservoir being infinite so that its temperature $T_2$ is
unchanged, as assumed above for simplicity. For a finite `reservoir', it and the system will reach an equilibrium temperature $T_3$, with $T_2<T_3<T_1$. 
If for a different cooling rate and different Joule heat generated the final equilibrium temperature were to be $T_4 \neq T_3$,
it would imply that either the system or the `reservoir' have negative heat capacity which is of course impossible. 
Because the system and the `reservoir' constitute our `universe', their final equilibrium temperature and their final states are
uniquely defined, and the considerations given above apply.

To resolve this inconsistency without violating well established laws of electromagnetism and thermodynamics, we would have to conclude
that the process happens infinitely slowly in nature, independent of the experimental conditions, e.g. the value of 
$\kappa$  in Fig. 3. That is contradicted by experiment. Alternatively, we would have to conclude 
that the final state of the superconductor is not unique, but depends on how the state was reached, i.e. fast or slowly. 
This would be a return to the pre-1933 view of superconductors. Since Gorter's and Casimir's work in 1934 \cite{gortercasimir}, continuing with
London's work and BCS, 
the premise that in a simply connected superconductor the state of the system for a given external magnetic field is unique has been
an essential  component of our understanding of superconductivity. Unless we want to abandon that cherished concept, we have
to conclude that the conventional theory is internally inconsistent, hence needs to be repaired or replaced.

  \section{thermodynamics of the phase transition}
  In this section we consider a somewhat  related problem, concerning the thermodynamics of the normal-superconductor transition. 
  We have considered some aspects of it already in previous work,  ref. \cite{entropy}. We will show here that
  the generation of Joule heat {\it does not} lead to difficulties in that case, in contrast to the   situation considered in the previous section.
  In the following section we explain the reason for the difference in the two situations.
  
         \begin{figure} [b]
 \resizebox{7.5cm}{!}{\includegraphics[width=6cm]{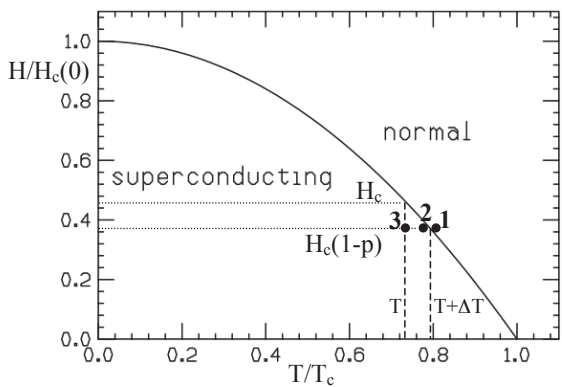}} 
 \caption {In the figure, state 1 corresponds to the normal state at the coexistence line and state 2 to the superconducting state at the coexistence line.
 State 3 is a state at a slightly lower temperature.
   }
 \label{figure1}
 \end{figure}

  Let us denote by $H_c(T)$ the critical field at temperature $T$ where normal and superconducting phases can coexist.
  Consider the three possible states shown in Fig. 4. In state 1, the system is in the normal state at temperature $T+\Delta T$.
In state 2, it is in the superconducting state at temperature  $T+\Delta T$, in state 3 it is in the superconducting state at temperature $T$.
The critical field at temperature $T$ is denoted by $H_c\equiv H_c(T)$, at temperature $T+\Delta T$ the critical field is
$H_c(1-p)\equiv H_c(T+\Delta T)$.

We consider the two possible routes for the transition shown in Fig. 5, in an applied  magnetic field $H_c(T+\Delta T)=H_c(1-p)$. In route A, the system undergoes the
N-S  transition while on the coexistence curve, at temperature $T+\Delta T$. The transition proceeds infinitely slowly and no Joule heat is generated. After the system reaches  the superconducting state, heat flows to the reservoir and the system cools to temperature $T$. 
In route B, the system in the normal state cools to temperature $T$ first, and then makes the  transition to the superconducting state.
Let us call $L(T)$ the latent heat for our system, i.e. the heat transferred out of the system when it goes from normal to superconducting at temperature $T$.

\subsection{Route A}
The transition proceeds infinitely slowly, since the system is on the coexistence curve (state 1 to state 2 in Fig. 4). 
Therefore, no Joule heat is generated. The total heat transferred from the system
to the heat reservoir between initial and final states is
\beq
Q_A=L(T+\Delta T)+C_s\Delta T
\eeq
to first order in $\Delta T$. We ignore the small change in $C_s$, the heat capacity in the superconducting state,  as the temperature changes between $T+\Delta T$ and $T$, because it is a second order
contribution. 

The latent heat is transferred between the system at temperature $T+\Delta T$ and the reservoir at temperature $T$. The change in the entropy of the universe in
this process is then
\beqn
\Delta S_{univ,A}&=&L(T+\Delta T)(\frac{1}{T}-\frac{1}{T+\Delta T}) \\ \nonumber
&+& O((\Delta T)^2) =  \frac{L(T)}{T}\frac{\Delta T}{T} +O((\Delta T)^2) .
\eeqn
The heat transfer during the process of temperature equalization generates no entropy to this order.

          \begin{figure} [t]
 \resizebox{7.5cm}{!}{\includegraphics[width=6cm]{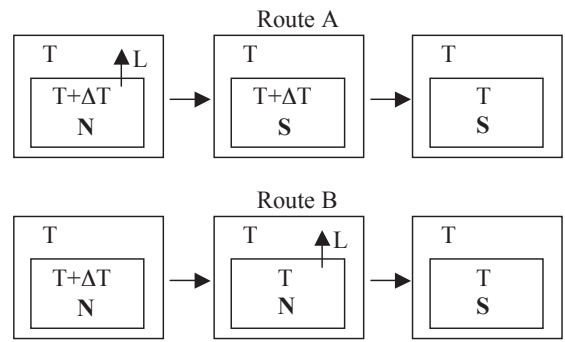}} 
 \caption {Two routes for normal-superconductor (N-S) transition in an applied magnetic field $H_c(T+\Delta T)$
 for a system initially at temperature $T+\Delta T$ that is put in thermal contact with a heat reservoir at temperature $T$.
 In route A, the system undergoes the transition to the superconducting state while at $T+\Delta T$, then cools. In route
 B, the normal system is cooled to temperature $T$, then it undergoes the transition to the superconducting state. L is the latent heat.
   }
 \label{figure1}
 \end{figure}

\subsection{Route B}
Route B is slightly more complicated. The system is cooled to temperature $T$ while still in the normal state, i.e. it is supercooled. Then it becomes superconducting by expelling the
applied magnetic field $H_c(1-p)$, which is smaller than the coexistence field $H_c=H_c(T)$ ($p>0$). This takes a finite amount of time and generates
Joule heat, because the changing magnetic flux in the normal region generates a Faraday field and a normal current. 
The total heat transferred to the heat reservoir here is
\beq
Q_B=C_n\Delta T+L(T)+Q_J
\eeq
where $Q_J$ is the Joule heat. The change in entropy of the universe in this process is
\beq
\Delta S_{univ,B}=O((\Delta T)^2) + 0 +\frac{Q_J}{T}  .
\eeq
In this equation, the first term corresponds to cooling in the normal state, which gives a second order contribution to the entropy; the second term is 
the change in the entropy of the universe when the latent heat and the Joule heat are  transferred between the system and the environment at the same temperature $T$, hence
it is zero. The third term, entropy generated by the production of Joule heat, accounts for the entire entropy generation to lowest order in $\Delta T$ in this process.
  
   Now the initial and final states are the same in both routes. Therefore we {\it must have}:
\bmath
\beq
\Delta S_{univ,B}=\Delta S_{univ,A}
\eeq
\beq
Q_B=Q_A
\eeq
\emath
     Is that so?
     
     For Eq. (71a) to be valid, we need, from Eqs. (68) and (70)
     \beq
     Q_J=\frac{L(T)}{T}\Delta T
     \eeq
  and for Eq. (71b) to be valid we need, from Eqs. (67) and (69)
     \beq
   C_n\Delta T+L(T)+Q_J=L(T+\Delta T)+C_s\Delta T .
   \eeq
   Assuming Eq. (72) does hold, Eq. (73) is to lowest order in $\Delta T$
   \beq
   C_s-C_n=-\frac{\partial L(T)}{\partial T}+\frac{L(T)}{T} .
   \eeq
   
   Now the Clausius-Clapeyron equation for a system with thermodynamic variables $T$, $V$, $P$ ($V$=volume, $P$=pressure) undergoing a first order phase transformation is well known:
   \beq
   \frac{dP}{dT}=\frac{L(T)}{T\Delta V}
   \eeq
   where $L$ is the latent heat and $\Delta V$ the volume change. The analogous equation for a superconductor is \cite{reif}
   \beq
   \frac{dH_c}{dT}=\frac{L(T)}{T(M_s-M_n)}
   \eeq
   where $M_n=0$ is the magnetization in the normal state and $M_s=-H_c/(4\pi)$ is the magnetization in the superconducting state, from which it follows that
   \beq
   \frac{L(T)}{T}=-\frac{H_c}{4\pi}\frac{dH_c}{dT}
   \eeq
   Replacing Eq. (77) in Eq. (74) we find
   \beq
   C_s-C_n=\frac{T}{4\pi}[(\frac{\partial H_c}{\partial T})^2+H_c \frac{\partial ^2H_c}{\partial T^2}]
   \eeq
   which is a well known relation, it follows from the formula
   \beq
   S_n(T)-S_s(T)=\frac{L(T)}{T}
   \eeq
   for the entropy difference, and $C_a=T(\partial S_a/\partial T)$ with $a=s$ or $n$. At $T_c$, the second term on the right-hand side
   of Eq. (78) is zero, and Eq. (78) reduces to the even better known Rutgers relation for the specific heat
   jump at $T_c$.
   
     Finally, we need to prove that the Joule heat dissipated in route B is indeed given by Eq. (72). 
     The calculation is similar to the one we did in ref. \cite{entropy} for the reverse process,
     the superconductor-normal transition. We discuss it in Appendix A, where we show that 
          \beq
     Q_J=\frac{H_c^2}{4\pi} p .
     \eeq
From $H_c (T+\Delta T)=H_c(T)(1-p)$
     we have
     \beq
     p=-\frac{1}{H_c}\frac{\partial H_c}{\partial T} \Delta T
     \eeq
     hence from Eqs. (80) and (81) and using Eq. (77)
     \beq
     Q_J=-\frac{H_c}{4\pi}\frac{\partial H_c}{\partial T}=\frac{L(T)}{T} \Delta T
     \eeq
     in agreement with Eq. (72).
     
     This also implies that there is no further entropy generation in the process of transferring momentum to the body in the process of 
     normal-superfluid conversion. However, we showed in Ref. \cite{entropy} that within the conventional theory this would not be possible:
     the process generates a momentum imbalance in the normal electron distribution \cite{halperin} that can only be resolved by transferring momentum
     to the body by normal scattering, thus generating additional Joule heat  \cite{entropy}. Therefore, already in Ref. \cite{entropy} we had concluded that
     the conventional theory violates the laws of thermodynamics.
     
     The two routes that we have considered here, A and B, correspond to having the situations in Fig. 3 where
     (A) $\kappa \rightarrow 0$ and (B) $\kappa \rightarrow \infty $, i.e.  heat being transferred from the system to the reservoir infinitely slowly and
     infinitely fast. Since we get the same result in the two limits, one without Joule heat, one with Joule heat, it is reasonable to assume we would get the same result  for
     any value of $\kappa$ and intermediate values for Joule heat. No inconsistency here.
  \newline
  \newline
     \section{similarities and differences between the processes in Sects. VIII and IX}
     
     In both Sect. VIII and Sect. IX we have considered processes where a magnetic field is expelled from the interior of a superconductor, 
     which according to Maxwell's electromagnetism generates a Faraday electric field. In both Sects. VIII and IX we have made 
     the usual assumption that normal electrons respond to an   electric field by generating a normal current,
     that undergoes normal scattering processes and generates Joule heat. Yet we have reached very different conclusions, namely that
     Sect. IX satisfies thermodynamics and Sect. VIII violates it. Let us compare the two situations.
     
  An  important difference is that in Sect. IX there is a natural mechanism that determines the speed at which the transition occurs in route B, 
     first elucidated by Pippard \cite{pippard}: as the magnetic field $H_c(1-p)$ is expelled at temperature $T$, the induced normal current generates a magnetic field in the same direction as the applied one,  that
     increases the magnetic field at the
     phase boundary to exactly $H_c=H_c(T)$. The transition cannot proceed faster because that would make the normal current larger and the  magnetic field at the phase boundary larger
     than $H_c$, reversing the direction of phase boundary motion. 
     It will also not proceed slower because microscopic times governing the normal-superconductor transition are obviously very  fast, and it is energetically favorable for the
     system to go superconducting as fast as it can. Therefore, the magnitude of $p$ sets the rate at which the transition will take place, and hence the amount of Joule heat
     that will be generated. There is no wiggle room.
     
     In contrast, there is no similar  mechanism in Sect. VIII to limit the speed of the transition. First, we can assume   that the entire process occurs at temperatures
     $T<<T(H_0)$, with $H_0=H_c(T)$, in other words, far from the coexistence curve (see Fig. 1). Therefore, the magnetic field generated by the normal current will not increase the magnetic
     field above the critical field at that temperature, in contrast to the situation in Sect. IX.  More importantly, in Sect. VIII the magnetic field generated by the normal current 
     will immediately induce (through the Faraday field created by it) a counter-supercurrent that will keep the magnetic field in the deep interior zero at all times. 
     Therefore, there does not appear to be any constraint on what the rate of the transition can be in Sect. VIII, other than the thermal conductivity of the
     heat conductor between the system and the reservoir. This leads to the absurd conclusion that in the limit where the heat conductor has arbitrarily large thermal
     conductivity the Joule heat generated becomes arbitrarily large. But there is no source of energy to generate such Joule heat.
     
     The essential  difference between the situations in Sects. VIII and IX is  that in Sect. IX the normal current was   induced in the normal region, and in Sect. VIII it was induced
     in the superconducting region.
     As a matter of fact, we  showed in ref. \cite{entropy} that the Joule heat Eq. (80) generated during the transition
     results from action of the electric field {\it in the normal region only}
     (Eqs. 29) to (36) of \cite{entropy}).
     In other words, even if not explicitly stated in ref. \cite{entropy}, we implicitly assumed, without justifying it,  that there was no Joule heat generated in the superconducting region
     within $\lambda_L$ of the phase boundary during the transition, even though an electric field does exist in that region.   If we had included such contribution, we would have obtained 
     a correction to Eq. (80) that would have spoiled 
      the agreement with thermodynamics. In light of this it is not surprising that we find disagreement with thermodynamics in Sect. VIII, where the
      Faraday field acts only in the superconducting region since there is no normal region. 
     
     The bottom line is: in order to get agreement with thermodynamics it is necessary to assume that an electric field does not induce a normal current
     in a superconductor at finite temperatures, where normal electrons and superfluid electrons coexist, when the 
     temperature changes. This is contrary to what the conventional theory of 
     superconductivity predicts.

     \section{Electromagnetic energy}
     It is interesting to analyze the allocation of electromagnetic energy in the process we are considering. According to
     Poynting's theorem (also used in Appendix A),
     \beq
     \frac{\partial }{\partial t}(\frac{B^2}{8\pi}) = -\vec{J}\cdot \vec{E} - \frac{c}{4\pi} \vec{\nabla}\cdot (\vec{E}\times \vec{B})
\eeq
where the left side is the change in electromagnetic energy density, the first term on the right side is 
minus the work done by the electromagnetic field on charges, and the second is minus the outflow of
electromagnetic energy. 

              \begin{figure} [t]
 \resizebox{6.5cm}{!}{\includegraphics[width=6cm]{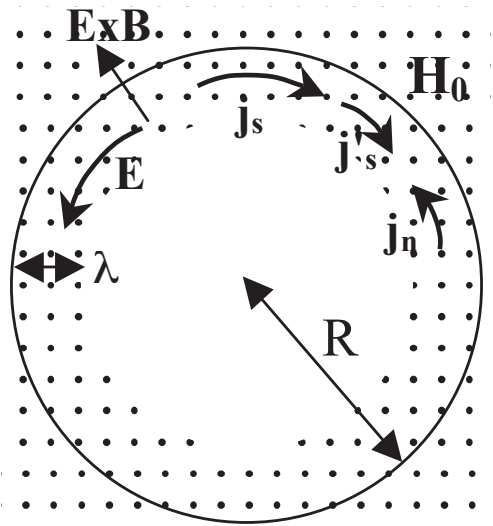}} 
 \caption {   Direction of electric field $\vec{E}$, supercurrents $\vec{j}_s$ and $\vec{j}'_s$, normal
 current $\vec{j}_n$ and Poynting vector (proportional to $\vec{E}\times\vec{B}$) in the process of $cooling$ the superconductor.  }
 \label{figure1}
 \end{figure}

Let us first assume there is no normal current. So Eq. (83) is 
     \beq
     \frac{\partial }{\partial t}(\frac{B^2}{8\pi}) = -\vec{j}_s\cdot \vec{E} - \frac{c}{4\pi} \vec{\nabla}\cdot (\vec{E}\times \vec{B})
\eeq
with $\vec{B}$, $\vec{E}$ and $\vec{j}_s$   given by eqs. (26), (28) and (29). 
The left side of Eq. (84)  is 
    \beq
     \frac{\partial}{\partial t}(\frac{B^2}{8\pi}) =\frac{H_0^2}{4\pi} \frac{R-r}{\lambda_L^2}
     e^{2(r-R)/\lambda_L}\frac{\partial \lambda_L}{\partial t}
     \eeq
     and the terms on the right side,  
     \beq
     -\vec{j}_s\cdot\vec{E}=-\frac{H_0^2}{4\pi \lambda_L}(1+\frac{R-r}{\lambda_L} )
     e^{2(r-R)/\lambda_L} \frac{\partial \lambda_L}{\partial t}
\eeq
and
\beqn
&-&\frac{c}{4\pi} \vec{\nabla}\cdot (\vec{E}\times \vec{B})= \\ \noindent
& &\frac{H_0^2}{4\pi}\frac{\partial }{\partial r} [ (1+\frac{R-r}{\lambda_L})   e^{2(r-R)/\lambda_L} ]  \frac{\partial \lambda_L}{\partial t}
\eeqn
and it is easily seen by substitution that Eq. (84) holds. 

In Eq. (84), the left side is negative (recall that $\partial \lambda_L/\partial t$ is negative), since the magnetic energy is decreasing as we cool and magnetic field lines move out.
The second term on the right is even more negative, more electromagnetic energy is flowing out than the decrease in
magnetic energy in the interior. This is because the first term on the right is positive,  the electric field decelerates the supercurrent,
hence does negative work on the charges. In other words, the supercurrent is giving energy to the electromagnetic field, energy that it acquired in the process of normal electrons condensing into the superfluid.

Assume now there is also normal current. At first sight it may seem that Poynting's theorem will be violated  
if we substitute $\vec{j}_s$ by $\vec{j}_s+\vec{j}_n$ in Eq. (84). The reason it is not is that when there is normal current
there is also an additional supercurrent, $\vec{j}'_s=-\vec{j}_n$. This new supercurrent is induced because the
normal current creates a magnetic field, and in that process Faraday's law creates a magnetic field that induces
this new supercurrent, which ensures that the magnetic field doesn't change, and in particular remains  zero in the
deep interior of the cylinder. So the energy balance equation is now
    \beq
     \frac{\partial }{\partial t}(\frac{B^2}{8\pi}) = -(\vec{j}_s+\vec{j}'_s)\cdot \vec{E}  -\vec{j}_n\cdot \vec{E} - 
      \frac{c}{4\pi} \vec{\nabla}\cdot (\vec{E}\times \vec{B})  
\eeq
with the same $\vec{B}$ and $E$ as before.
So Poynting's theorem continues to hold, but now there are two terms in Eq. (89) representing work done by the
electromagnetic field on charges. The first term is positive and larger than before, since the supercurrent is
larger, so the field is doing more negative work on the supercurrent than before. The second term is
negative, representing the positive work that the field does on normal charges, since the normal current is in the
same direction as the electric field, and that work is dissipated as
Joule heat. Figure 6 shows the directions of currents, Faraday field and Poynting vector in the cooling process.

The faster the process goes, i.e. the larger the thermal conductivity $\kappa$ of the heat conductor in Fig. 3 is,
the larger will be $j'_s$, $j_n$, and the associated Joule heat $Q_J$ and associated 
Joule entropy $S_J$. `Somebody' has to supply the energy to create $j'_s$, that is transferred to   $j_n$
through the electromagnetic field and then converted into Joule heat.   The only  `somebody' here   is the superconductor.

As the superconductor cools,  electrons condense into the
superfluid and thereby lower their energy. Part of that energy difference is  heat transferred to the reservoir, and part is transferred
to the electromagnetic field via the first term on the right in Eq. (84). When the temperature decreases from $T$ to $T-\Delta T$
we can write for these terms
\beq
\Delta E=\Delta Q+\Delta E_{em} .
\eeq
Both $\Delta E$ and $\Delta Q$ are determined by the initial and final temperatures, $T$ to $T-\Delta T$. In particular,
$\Delta Q=T(\partial S/\partial T)\Delta T$. So $\Delta E_{em}$, the energy transferred to the electromagnetic field, is fixed.
It is given by the space and time integral of the second term in Eq. (84), which yields
\beq
\int _0^\infty dt \int d^3 r (-\vec{j}_s\cdot\vec{E})=\frac{3}{8} H_0^2 R h [\lambda_L (T)-\lambda_L(T-\Delta T)]
\eeq
and is independent of how fast or slow the cooling is. Therefore, there is no extra energy to create $j'_s$,
therefore it is not possible that $j_n$ exists either.  

     \section{more on thermodynamics}
     We have argued that generation of Joule heat in a process where the temperature of the superconductor changes, always below $T_c$, leads to violation
     of both the first and second law of thermodynamics. In this section we consider the possibility that through some unknown mechanism
     conservation of energy can be maintained, and show that even in that case the second law would be violated.
     
     Consider an intermediate step in the process, where the system changes its temperature from $T$ to 
$T-\Delta T$. Consider two different ways to do this step:

(a) Infinitely slowly

(b) In a finite amount of time, $\Delta t$.

\noindent According to the previous discussion, for (b) finite  Joule heat will be generated in the
system.

The  total amount of heat transferred from the system to the reservoir  has to be the same for (a) and (b), since the initial and final states of both
the system and the electromagnetic field are the same, and hence also those of the reservoir. Let's call that heat $\Delta Q$. 
For process (a), we have 
\beq
\Delta Q=C(T) \Delta T
\eeq
where $C(T)$ is the equilibrium heat capacity of the system. No Joule heat is generated.

For process (b), assume Joule heat $\Delta Q_J$ is generated. One could imagine that the heat capacity of the system is
different than the equilibrium one when the process occurs at a finite rate and  involves Joule heat, let's call it $C_r(T)<C(T)$. We will  then have for process (b)
\beq
\Delta Q=C_r(T) \Delta T+\Delta Q_J  
\eeq
transferred from the system to the reservoir, the same as in process (a), respecting the first law. Both terms in Eq. (93)  would depend on the speed of the process, the faster the process the smaller the first term and the larger the second term.

However,  consider the change in entropy of the universe. The change in entropy of the reservoir in both processes is
\beq
\Delta S_{res} =\frac{\Delta Q}{T_2}
\eeq
and the change in entropy of the system in both processes is (to lowest order in $\Delta T$)
\beq
\Delta S_{sys} =-\frac{\Delta Q}{T}
\eeq
In process (a) that is all there is, so the change in entropy of the universe   is
\beq
\Delta S_{univ}^{(a)}=  \Delta S_{res}+ \Delta S_{sys}=   \frac{\Delta Q}{T_2}-\frac{\Delta Q}{T}  
\eeq
which is of course larger than zero since $T>T_2$. 
In process (b), in addition to these, we need to take into account that  generation of Joule heat generates entropy:
\beq
\Delta S_J=\frac{\Delta Q_J}{T} .
\eeq
 Therefore, the change in entropy of the universe in process (b) is
\beq
\Delta S_{univ}^{(b)}=\Delta S_{univ}^{(a)} + \Delta S_J >  \Delta S_{univ}^{(a)}.
\eeq
However, entropy is a function of state. Therefore, the second law of thermodynamics is violated
by Eq. (98).

              \begin{figure} [t]
 \resizebox{7.5cm}{!}{\includegraphics[width=6cm]{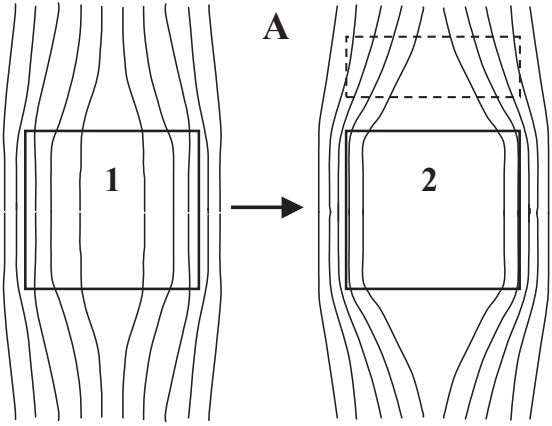}} 
 \caption {Schematic depiction of field lines in a cylindrical superconductor of finite height at
 a higher (left panel) and a lower (right panel) temperature.     }
 \label{figure1}
 \end{figure}

Finally, to try to get around this problem let us consider the possibility that the system lowers its entropy by a larger amount than Eq. (95)  in the process
of transferring heat to the reservoir in process (b). This would happen if the system is at a slightly lower temperature. Indeed one can argue that
to generate the normal current the system has to supply energy, and that would lower its temperature from $T$ to $T-\delta T$,
where $\delta T$ is determined by the equation  
\beq
\Delta Q_J=C(T)\delta T .
\eeq
The change in entropy of the system would then be, rather than Eq. (95)
\beq
\Delta S_{sys}^{(b)} =-\frac{\Delta Q}{T-\delta T}=-\frac{\Delta Q}{T}-\frac{\Delta Q}{T}\frac{\delta T}{T} 
\eeq
or, using Eqs. (99) and (92)
\beq
\Delta S_{sys}^{(b)}  =-\frac{\Delta Q}{T}-\frac{\Delta Q_J}{T}\frac{\Delta T}{T} 
\eeq
so that the change in entropy of the universe in process (b) would be
\beq
\Delta S_{univ}^{(b)}=\Delta S_{univ}^{(a)}  + \Delta S_J (1-\frac{\Delta T}{T} )
\eeq
and the violation of the second law of thermodynamics is not resolved.

     \section{an analogy}
     To understand the problems encountered with thermodynamics  better we will consider an analogy here.
     Figure 7 shows schematically field lines for a superconductor of finite height in a uniform magnetic field.
     As the temperature is lowered from $T_1$ (left panel) to $T_2$ (right panel) the   London
     penetration depth decreases  and the magnetic field lines move outward.
     The initial and final states correspond to points 1 and 2 in the phase diagram of Fig. 1, and are the same no matter
     how fast or slow the process of cooling the superconductor is.

   \begin{figure} [t]
 \resizebox{7.5cm}{!}{\includegraphics[width=6cm]{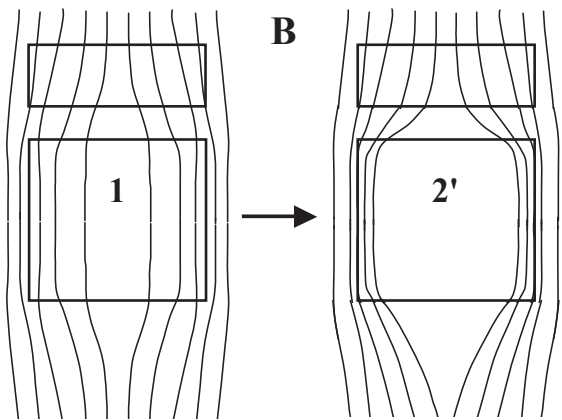}} 
 \caption {Same superconducting cylinder as in Fig. 7, with a normal metal cylinder added on top (smaller rectangle)
 The figure shows schematically the situation where the normal metal is a perfect conductor, or equivalently the 
 transition occurs infinitely fast, so the field lines are frozen in the normal metal.
   }
 \label{figure1}
 \end{figure}

 Imagine now that we place a small normal metal cylinder on top of our superconducting cylinder, as shown in 
 Fig. 8, and we repeat the experiment. The final state will depend on the electrical conductivity
 of the normal metal and the speed at which the transition occurs. In particular, the final state 
 will be different if (A) the transition occurs infinitely slowly, or equivalently if it occurs at any rate with the normal material being 
 insulating rather than metallic, or (B) infinitely fast, or equivalently at any rate with the normal material being a
 perfect conductor.
 
 Namely, in case (A) the presence of the normal cylinder will not have any effect, and the final state will be the same
 as depicted in Fig. 7, right panel (the normal metal cylinder is indicated as a dashed rectangle in Fig. 7).

 Instead, in case (B), the initial field lines will be $frozen$ in the normal metal cylinder. When the superconductor is cooled
 it will  expel the magnetic field lines to a penetration depth $\lambda_2 < \lambda_1$ but it will have
 a harder time doing so, because of the boundary condition that magnetic field lines in the normal metal cannot move.
 As a consequence, the final state in Fig. 8 for the superconductor, denoted by 2',  will not be the same as in Fig. 7.
 In particular, the magnetic field will be more intense near the surface than in the case of Fig. 7.
 The magnetic field lines will be somewhat different  and the  associated  supercurrent will also be somewhat different.
 
 Is it possible to have two different final states for a simply connected superconductor? Yes, because the boundary
 conditions are different, the magnetic field outside the superconductor near the upper surface is different for the two situations
 depicted in Figs. 7 and 8.  
 
 More generally, if the normal metal cylinder has a finite electrical conductivity and the cooling process occurs at a finite rate, 
 Joule heat will be dissipated in an amount that depends on the conductivity and the rate. For each different case,
 with different Joule heat dissipated, the final state of the superconductor will be different. If we imagine this system
 in the setup of figure 3, where it dumps heat into a reservoir at temperature $T_2$, the heat dumped will
 be different in each case and the total entropy generated will be different in each case.
 There is no contradiction with thermodynamics here, because both the system and the reservoir are  reaching a different final state in each case.
 If we were to compute all the thermodynamics quantities we would find that the laws of thermodynamics hold.

 I argue that in a sense what Fig. 8 depicts is the view of superconductors within the conventional theory of
 superconductivity, where
 the `normal metallic cylinder' is immersed inside the superconductor rather than outside as in Fig. 8.
 Thermodynamic laws will be satisfied  $if$ the system is allowed to reach different final states depending on the
 speed at which the cooling occurs and of the magnitude of its   electrical conductivity in the normal state. For example, if the
 cooling is fast the system will ``use up'' a large part of its condensation energy in Joule heat rather than in expelling magnetic
 field, and the final state will have a larger London penetration depth. 
 However, the conventional
 theory of superconductivity says that there is a unique final state. Hence it is in conflict with thermodynamics.
 
  \section{further arguments}
  Some readers may think that the situation considered here is not very  different from  other situations where Joule heat is generated
  in an amount that depends on the speed of the process, hence that the arguments given here would imply inconsistencies in other situations, hence  these arguments
  cannot be right. That is not so. Let us analyze a couple of simple examples and how they differ from the situation considered here.
  
  \subsection{Magnet and normal conductor}
  Consider a situation where we approach a permanent magnet by hand to a normal metal. Eddy currents are generated and Joule heat
  is dissipated, in different amounts depending on the speed of the process and the conductivity of the material.
  In particular, if we approach the magnet faster the Faraday field is larger and more Joule heat is dissipated. Why is thermodynamics not
  violated?
  
  The answer is of course that if we approach the magnet faster the eddy currents generate a larger magnetic field that generates a 
  larger force that opposes the motion of the magnet, hence our hand has to do more work to approach the magnet to the metal
  at a faster speed, and the extra work supplied by us supplies the extra Joule heat generated. Similarly if the conductivity of the
  material is larger the Joule heat generated is larger, but so are the eddy currents generated and the opposing force.
  
  Instead, for the case of the superconductor considered in this paper there is no external ``hand'' that can supply variable
  amounts of work depending on the speed of the process. As discussed, initial and final states of the metal and the reservoir, and hence
  their energies, are fixed by the initial and final temperatures and are independent of the speed of the process, but the
  speed of the process is not fixed (it depends on $\kappa$) and neither is the amount of 
  generated Joule heat according to the conventional theory.
  
    \subsection{Inductor in a circuit}
    Let us consider a circuit with an inductor $L$ in series with a resistor $R$ connected to a battery, through which a current circulates.   A magnetic field will exist in the inductor, 
    and the magnetic energy stored is
    \beq
    U=\frac{1}{2}LI_0^2
    \eeq
    with $I_0$ the initial current circulating. If we now disconnect the battery and short the circuit, the current will decay with time constant
    $t_0=L/R$, given by
    \beq
    I(t)=I_0e^{-t/t_0}
    \eeq
    generating power
    \beq
    P(t)=I(t)^2R.
    \eeq
    If $R$ is large the Joule heat generated per unit time is large, if $R$ is small it is small, why is energy conservation not violated?
    
    The answer is of course that the  $integrated$ power, that  gives the total Joule heat generated as the current decays to zero, is independent
    of $R$  and yields of course Eq. (103), as is easily verified.
 If $R$ is small, the process proceeds very slowly and the power dissipated at any given time is very small,
    if $R$ is large the process is fast and the power is large, but the time integral and total Joule heat dissipated is always
    the same, independent of the speed of the process. And of course the origin of the energy dissipated in Joule heat is
    accounted for, it is the energy that was originally stored in the inductor.
    
    Instead, for the case of the superconductor discussed here, the origin of the energy dissipated in Joule heat is not accounted for, the speed of the process is not determined by the conductivity of
    the normal electrons, nor does the magnetic field and its time variation depend solely on the normal current since there is also
    supercurrent. For those reasons the total Joule heat generated is not the same independent of the speed of the process, and is not
    the same independent of the normal resistivity, and in
    particular goes to zero if the temperature change proceeds infinitely slowly and/or if the normal state resistivity goes to infinity, and is not zero if it proceeds at a finite rate and the normal state resistivity is finite,
    leading to the conflict with thermodynamics discussed earlier in the paper.

 \section{resolution of the conundrum}
 In this section we propose a qualitative solution to the conundrum discussed in the previous sections.
 
 We don't want to give up on the principle that superconductivity is a thermodynamic state of matter, including
 the situation where an external magnetic field is present. Nor do we want to give up on the laws of
 thermodynamics, electromagnetism or mechanics. We propose that there is only one way to resolve this
 conundrum.

 Consider the situation in fig. 8. Imagine that there are electric charges in the normal metal that are completely free to move, 
 and the metal is a perfect conductor. If the mobile charges in the metal move together with the magnetic field lines as the
 system is cooled, no Joule heat will be generated. This is what happens in a perfectly conducting plasma according to
 Alfven's theorem \cite{alfventh}: the magnetic field lines are frozen into the fluid, when the fluid moves the magnetic field lines move
 with it and vice versa, and no Joule heat is generated.
 
 Therefore,   if a perfectly conducting fluid residing in the interior of the superconductor moves outward
 together with the field lines when the system is cooled, no Joule heat will be generated and the system will be able
 to reach a unique final state independent of the rate at which the field lines move out.
 
 For this to be a possibility   this  fluid has to be both charge neutral and mass neutral,
 so that neither charge nor mass accumulation near the surface will occur.  
 
This is  possible if there is an outward flow of both electrons and holes. Because holes are not real physical particles,
 the outward motion of holes is associated with inward rather than outward  flow of mass, and it can compensate the outflow of
 mass due to electrons. The electrons and holes flowing out can drag the magnetic field lines with them, as in a perfectly conducting
 plasma \cite{alfventh}, without energy dissipation. 
 
 In previous work, we have explained the dynamics of the Meissner effect within the theory of hole superconductivity as follows
 \cite{momentum,entropy,revers}:
 when the phase boundary moves outward, electrons becoming superconducting move outward and acquire azimuthal speed due to the
 action of the Lorentz force, giving rise to the increasing Meissner current.
 At the same time, normal holes move outward  to compensate for the radial charge imbalance, and the combined action of magnetic and electric fields
 cause the holes to move radially out without acquiring azimuthal velocity \cite{revers}. In this  process they transfer azimuthal momentum to the 
 body as a whole without any energy dissipation, thus compensating for the increasing azimuthal momentum of the Meissner current.
 
 We believe the same physics is at play here, as the system is cooled and normal electrons condense into the superfluid,
 and can explain the contradictions encountered within the conventional theory.
 In particular, normal holes moving out are subject to a clockwise Lorentz force which can exactly compensate for the 
 counterclockwise force exerted by the Faraday field on them, so that no azimuthal current and no Joule dissipation results.
 
 The details of the process will be discussed elsewhere.
 
 \section{discussion}
  
  In this paper we have analyzed the process where a type I superconductor in a magnetic field is cooled while always  in the
  superconducting state. We have encountered several problems in trying to understand this process within the conventional
  theory of superconductivity, and we have suggested that the alternative theory of hole superconductivity \cite{holesc} may be able
  to resolve these problems. The same problems would have been encountered if we had considered a process of heating rather
  than cooling.
  
  The first problem we pointed out is that there has to be a physical mechanism for electrons that go from normal to superconducting in the process
  of cooling to
  spontaneously acquire  angular momentum, and at the same time for the body to acquire equal and opposite angular momentum.
  The physical mechanisms by which this happens have not been explained in the literature on conventional superconductivity.
  We believe the conventional theory cannot explain these processes. In fact, the same question arises
  in the normal-superconductor phase transition in a magnetic field, or its reverse \cite{disapp}. We have analyzed 
  these questions  within the alternative theory of hole superconductivity and shown how they can be explained with physics that
  is not part of the conventional theory \cite{momentum,revers,disapp}. The same physics would explain these momentum changes in the context discussed in this paper.
  
  The next problems relate to the action of the Faraday electric field that is necessarily induced  when the temperature changes. 
  It will generate a normal current  and cause dissipation of  Joule heat. We showed that this Joule heat will become arbitrarily large sufficiently close to
  the phase transition point,  and that it will become arbitrarily large if the rate of temperature change is large. 
.
  
  We have shown that in fact there is  normal current originating from two different sources. On the one hand, the Faraday electric field
  gives rise to a normal current proportional to the density of normal electrons, just like in a normal metal.
  On the other hand, within the conventional theory the process of condensation leaves behind a momentum imbalance in the
  normal electron distribution \cite{halperin, entropy}, which gives rise to normal current {\it in the same direction} as that
  induced by the Faraday field, so it adds to it. The decay of this normal current necessarily gives rise to Joule heating during the process where
  the London penetration depth changes, either cooling or heating.
  
Note also that the amount of Joule heat $Q_J$ generated depends both on the speed of the process and on the normal
conductivity $\sigma_n$, two variables that are not tied to each other, contrary to what happens in the phase transition
(see Sect. IX and Refs. \cite{pippardmine,pippard}), and contrary  to what happens in a circuit with inductor and resistor (Sect. XIV B).
So for a given $Q_J$, we can get the same $Q_J$ by making the process slower by a factor of 2 and increasing
$\sigma_n$ by a factor of 2. Therefore the conflict with thermodynamics that  we encounter is not related to the degree
by which the system is `out of equilibrium' during the process.
  
We have found  that the generation of $any$ Joule heat in the process considered in this paper  is inconsistent with thermodynamics. The only way
  to make it consistent with thermodynamics within the conventional theory  would be to assume that the system reaches different final states depending on the speed of
  the process, which is itself inconsistent with the conventional theory. By contrast,
  we showed in this paper that the generation of Joule heat during the normal-superconductor transition is consistent with thermodynamics, assuming
  Joule heat is dissipated only in the normal region $and$ assuming that the process of normal/superfluid conversion is able to transfer momentum
  to the body without dissipation \cite{entropy}.
  
In the situation discussed in this paper, the electric field arises in a process
  where the superfluid density is changing. Therein lies the essential difference with the situations considered in ref. \cite{tinkhamac}
  within the conventional theory,
  where the electric field
  is due to an electromagnetic wave or an ac current, and is not directly associated with changes in the superfluid density.
  In that case, it is well established experimentally and theoretically that the action of the electric field on normal electrons gives rise
  to dissipation. Within the conventional theory of superconductivity the response of normal electrons in those  situations
  and in the situation considered here  will be the same. This is necessarily so within the conventional theory in order to satisfy momentum conservation.
But it  leads to the conundrum discussed in this paper.
  
  Within the conventional theory these normal currents and the resulting Joule heat are unavoidable for a simple reason. 
There is {\it no mechanism} that can transfer momentum between electrons and the body as a whole in the conventional theory that
does not involve scattering of normal electrons through the same processes that give rise to resistivity in the normal state.
Transferring momentum between electrons and the body is necessary to conserve momentum. As we have seen, the total momentum of
the body does not change in the process, but the Faraday field imparts the positive ions  momentum in the counterclockwise direction that needs to be
compensated by `somebody' giving the body  clockwise momentum. In addition, the condensation process involves normal electrons acquiring 
counterclockwise momentum, which needs to be compensated by the body acquiring more clockwise momentum. Both those processes 
necessitate transfer of momentum from normal electrons to ions through normal scattering processes in the conventional theory, that give rise to 
Joule heating as discussed in 
Sects. VI and VII and leads to the conflict with thermodynamics.

Instead, within the theory of hole superconductivity there is a mechanism to transfer momentum between electrons and 
the body that does not involve normal scattering processes and associated Joule heating. It requires that the normal
charge carriers are $holes$ \cite{momentum,revers,entropy,disapp,whyholes}.

We have pointed out that the theory of hole superconductivity \cite{holesc}  has physical elements not contained in the conventional theory that
may provide a way to resolve this conundrum. Within this theory the response of the superconductor to ac fields would be similar to the
conventional theory \cite{frankac}, consistent with experiment. However, the response to the electric field that arises in the process of
electrons condensing or decondensing into or out of the superfluid would not be the same. That is because within this theory 
those processes are associated with radial outflow and inflow of electrons and holes, and the magnetic Lorentz force acting on these moving
charges will change the effect of the Faraday field on them. In addition, the condensation process does not give rise to
normal current within this theory \cite{entropy}. Qualitatively, within this theory condensation and decondensation is associated
with radial flow of a charge-neutral mass-neutral perfectly conducting plasma, which according to Alfven's theorem
will flow without dissipation \cite{alfventh}.

 \appendix

\section{Calculation of the Joule heat for the normal-superconductor transition}

We consider the electromagnetic energy equation
\beq
\frac{d}{dt}(\frac{H^2}{8\pi}) = -\vec{J}\cdot \vec{E} - \frac{c}{4\pi} \vec{\nabla}\cdot (\vec{E}\times \vec{H}).
\eeq
 for the case where the system makes the transition from normal to superconducting in an applied
magnetic field $H_c(1-p)$. 
 The left side represents the change in energy of the electromagnetic field as the magnetic field  
 is expelled from  the body, the first term on the right side is the work done by the electromagnetic
 field on  currents in this process, and the second term is the outflow of electromagnetic energy. Integrating over the volume of the body $V$  and over time we find for the
 change in electromagnetic energy per unit volume
 \beq
\frac{1}{V} \int d^3r \int_0^\infty dt \frac{d}{dt}(\frac{H^2}{8\pi}) =-\frac{H_c^2 (1-p)^2}{8 \pi} .
 \eeq
 since at the end the initial magnetic field $H_c(1-p)$  is completely excluded from the body.
 From Faraday's law and assuming cylindrical symmetry we have for the electric field generated by the changing magnetic flux at the surface
 of the cylinder
 \beq
 \vec{E}(R,t)=-\frac{1}{2\pi R c}\frac{d}{dt}  \phi(t) \hat{\theta}
 \eeq
 where $R$ is the radius of the cylinder and $\phi(t)$ is the magnetic flux throught the cylinder, with
  $\phi(t=0)=\pi R^2 H_c (1-p)$, $\phi(t=\infty)=0$. 
Integration of the second term on the right in Eq. (A1), the energy outflow, over space and  time,
converting the volume integral to an integral over the surface   of the cylinder, 
using that $H=H_c(1-p)$ at the surface of the cylinder independent of time and Eq. (A3)
for the electric field at the surface yields
 \beq
 \frac{1}{V} \int_0^\infty dt    \oint (- \frac{c}{4\pi}) (\vec{E}\times \vec{H}) \cdot d\vec{S}=-\frac{H_c^2 (1-p)^2}{4 \pi} .
 \eeq
This gives the total electromagnetic energy flowing out through the surface of the sample during the
 transition.
 
 The current $\vec{J}$  in Eq. (A1) flows in the azimuthal direction and is given by the sum of superconducting and normal currents
 \beq
J(r)=J_s(r)+J_n(r)
\eeq
where $J_s(r)$ flows in the region $r\leq r_0(t)$ and is of appreciable magnitude only within $\lambda_L$ of the phase boundary, where $\lambda_L$
is the London penetration depth. $r_0(t)$ is the radius of the phase boundary  at time $t$. 
Integration of the second term in Eq. (A1) over the superconducting current yields \cite{pippardmine}
\beq
\frac{1}{V} \int d^3r \int_0^\infty dt  (-\vec{J}_s \cdot \vec{E} )=  \frac{H_c^2}{8\pi}  .
\eeq
This is because the Faraday field decelerates the supercurrent \cite{pippardmine} as the phase boundary
moves out .

The Joule heat per unit volume generated during the transition is
\beq
Q_J \equiv \frac{1}{V} \int d^3r \int_0^\infty dt  \vec{J}_n \cdot \vec{E}
\eeq
hence from integrating Eq. (A1) over space and time using Eqs.  (A2), (A4), (A5) and (A6) we have
\beq
-\frac{H_c^2 (1-p)^2}{8\pi}= \frac{H_c^2}{8\pi} - Q_J - \frac{H_c^2 (1-p)^2}{4\pi}
\eeq
which implies 
\beq
Q_J = \frac{H_c^2}{4\pi}p 
\eeq
to linear order in $p$. The entropy generated from Joule heat is then
\beq
\Delta S_{Joule}=\frac{Q_J}{T}= \frac{H_c^2}{4\pi T} p  .
\eeq
The reason for why in this case there is `extra' condensation energy to generate the Joule heat
is that the system is expelling magnetic field $H_c(1-p)$ while being  at a temperature where 
its critical field is larger, i.e. $H_c$, and correspondingly it has the necessary extra condensation energy.
 
It is interesting that in this calculation we have not made an assumption of what the speed of the process is, and yet the amount
of Joule heat and Joule entropy are completely determined. The reason is, the conditions of the problem completely determine
what the speed of the process is. In ref. \cite{entropy} we showed that if we calculate explicitly the Joule heat 
from Eq. (A7) over the time the process takes, instead of obtaining $Q_J$ from Eq. (A8) by substraction, the same answer
Eq. (A9) is obtained.
\newline

\noindent {\bf Note added:}
This paper was submitted for publication in Physical Review B and rejected on the advice of three referees. The interested
reader can find  the referees' arguments and the author's responses here
 \footnote{See Appendix B of the paper available  at 
 \href{https://jorge.physics.ucsd.edu/inconsistency}{https://jorge.physics.ucsd.edu/inconsistency}
 }.

  \acknowledgements
 The author is greatly indebted to Tony Leggett for motivating him to consider this problem and for  extensive 
 and stimulating discussions on it and related problems.  He is also grateful to Bert  Halperin for extensive and  stimulating discussions on this and related problems.
    
    \end{document}